\documentclass[a4paper,8pt]{article}
\usepackage{geometry}
\usepackage[utf8]{inputenc}
\usepackage{amsmath}
\usepackage{amssymb}
\usepackage{appendix}
\usepackage{latexsym}
\usepackage{breqn}
\usepackage{graphicx}

\title{Creating spacetime shortcuts with gravitational waveforms}
\author{Charles J. Quarra}

\begin{document}

\maketitle

\begin{abstract}
A region-delimited gravitational wave field can be constructed, such that a subset of geodesics crossing this region will move faster than nearby geodesics moving entirely inside flat spacetime, along a preferred direction. Null geodesics inside this region will move faster-than-light according to far away observers. The waveform is synthesized from homogeneous plane wave solutions, and the resulting field is the gravitational equivalent of a Gaussian beam.
\end{abstract}

Several authors \cite{Alcubierre:1994tu, Krasnikov:1995ad, Natario:2001tk} have proposed mechanisms within the standard theory of General Relativity (GR) to allow some level of circumvention around the light speed limit, by warping the spacetime geometry in some compact region. However, all the mechanisms proposed so far require the engineered spacetime region to be filled with matter that violates well established energy conditions, and is not known to exist in nature\cite{Ford:1994bj}. But even ignoring the problem of violation of the energy conditions, these geometries have other problems related to acausal setup of the exotic matter distribution\cite{Coule_No_warp_drive_98}, as well as quantum instabilities in the semiclassical limit\cite{Finazzi:2009jb}

However, the idea of using matter to curve surrounding spacetime does not exhaust the possibilities that GR offers in order to create customized geometries. Gravitational waves (GW) are themselves perturbations of geometry that travel at the speed of light. Even while the full theory of GR is a nonlinear theory, the principle of superposition still applies within the limit of weak plane waves, and one can consider some superpositions of such planar waves physically valid perturbations.  The present work shows that for specially crafted gravitational waveforms of this type, one can produce geometries in pure vacuum with Faster-Than-Light (FTL) properties, similar to those obtained via other geometrical drives.

We illustrate in this work that the geometry of null congruences can be affe\-cted in a way that allows FTL communication. To be precise, we construct a focal region of a gravitational waveform composed of traceless and transverse planar waves, and find that null congruences entering the focal region can become asymptotically accelerated, such that they arrive effectively before similar geodesics that do not enter the field region, according to distant observers. The asymptotic delay or advancement of congruences will be affected by the local phase of the perturbation at the moment the geodesic enters the region, the period of the oscillation, as well as the width of the focal region.

\section*{Gaussian geometry}
\label{eq:gaussian_geometry_section}
Our geometry of study is obtained by considering the gravitational equivalent of a Gaussian beam. Gaussian beams are one of the most basic propagating fields used in optical applications, and its general properties are inherited from the wave equation (and corresponding Helmholtz equation). However, there are differences. Optical fields are oscillations of the electromagnetic vector field $A_{\mu}$, while gravitational fields are tensor perturbations $h_{\mu \nu}$ with two physical degrees of freedom for each mode, so is not an straightforward realisation that there should be a simple equivalent in the gravitational case.

In linearized gravity, given appropiate harmonic coordinates, homogeneous solutions of the Einstein equations in vacuum propagate via the standard wave equation

$$ \Box h^{TT}_{\mu \nu} = 0 $$

Where the perturbation $h^{TT}$ is a purely transverse and traceless perturbation, as our goal is to build a field composed only from homogeneous solutions to the gravitational wave equation. In order to begin our construction, we start with a simple plane wave solution in the TT-gauge

$$D_{\mu \nu} e^{-I k_{\alpha} x^{\alpha}} $$

With $D_{\mu 0}=0$, as well the traceless $D_{i i}=0$ and transverse $k_i D_{i j}=0$ conditions. We now consider a generic plane wave component with a polarization $D_{\mu \nu}(k)$, to denote the fact that the polarization of each plane wave depends only on the vector $k_{\alpha}$. Once the functions $D_{\mu \nu}(k)$ are given, we can make a Fourier integral on each component to obtain the components of the physical metric perturbation.

Transverse conditions results in 3 equations given by

$$k^i D_{ij}(k) = 0 $$

As well as an additional equation for the traceless condition 

$$D_{ii}(k)=0$$

This results in a system of equations. We also replace $k_z = \sqrt{k^2 - k_x^2 - k_y^2}$ with $k= \omega/c$, and we make further choices for the polarization of the perturbation:

\begin{align*}
D_{xx}(k) &= h_{+} \\
D_{xy}(k) &= D_{yx}(k) = 0
\end{align*}

In our analysis $h_{\times}$ has been set to zero for simplicity. This gives us enough information to solve for all the $D_{ij}(k)$ functions

\begin{equation}
D_{\mu \nu}(k)=
\begin{bmatrix}
0 & 0 & 0 & 0 \\
0 & h_{+} & 0 & - \frac{ k_x h_{+}}{\sqrt{ \omega^2 /c^2 - (k_x^2 + k_y^2)  }} \\
0 & 0 & - \frac{h_{+} (\omega^2 - k_y^2 c^2) }{\omega^2 -k_x^2 c^2} & \frac{h_{+} k_y (\omega^2 - k_y^2 c^2) }{(\omega^2 -k_x^2 c^2) \sqrt{\omega^2 /c^2 - (k_x^2 + k_y^2) } } \\
0 & - \frac{ k_x h_{+}}{\sqrt{ \omega^2 /c^2 - (k_x^2 + k_y^2)  }} & \frac{h_{+} k_y (\omega^2 - k_y^2 c^2) }{(\omega^2 -k_x^2 c^2) \sqrt{\omega^2 /c^2 - (k_x^2 + k_y^2) } } & - \frac{h_{+} c^2 (k_y^2 - k_x^2) }{\omega^2 -k_x^2 c^2} 
\end{bmatrix}
\end{equation}

for all possible null wavevectors $k^{\mu}$ with $k_{\mu} k^{\mu}=0$ and $k_z > 0$, the corresponding plane wave component of the perturbation is a well-defined finite amplitude. We now multiply our polarization functions by the wavevector gaussian envelope

$$ \frac{ \sigma^2}{2} e^{ -(k_x^2 + k_y^2)(\sigma^2/4 + I c z / 2 \omega ) } e^{ I (k z - \omega t) } $$

and perform a 2D Fourier transform on the $xy$ plane to obtain the metric perturbation

\begin{equation}
 \label{eq:htt_transform}
 h^{TT}_{\mu \nu}= \frac{\sigma^2}{2} e^{ I (k z - \omega t) } \int_{\mathbb{R}^2} e^{ -(k_x^2 + k_y^2)(\sigma^2/4 + I c z/ 2 \omega ) } D_{\mu \nu}(k) e^{ -I(k_x x + k_y y) } dk_x dk_y
\end{equation}

Most component transforms are nontrivial to compute, but as becomes clear from eqs. \ref{eq:geodesic_metric_components} in the Appendix, the only component needed for the purpose of this work is the $xx$ component:

\begin{equation}
\label{eq:xx_component}
 h^{TT}_{xx}= \text{Re} \left[  \frac{ h_{+} \sigma^2 e^{ I (k z - \omega t) } }{\sigma^2 +  2 I c z / \omega } \exp{\left( - \frac{x^2 + y^2}{\sigma^2 + 2 I c z/ \omega} \right)} \right]
\end{equation}

We will focus on the behavior of geodesics crossing \emph{transversally} this gravi\-tational Gaussian beam in the field region near the waist at $z=0$. We will denote this field region $\mathcal{R}$. We assume that the Gaussian beam propagates in otherwise flat Minkowsky spacetime.

The goal will be to observe and understand the asymptotic delay of geodesics travelling along the $x$-axis inside $\mathcal{R}$. This asymptotic delay is to be understood not only compared to its expected value in a identical spacetime with no gravitational waves, but also compared to other geodesics that do not cross the focal region $\mathcal{R}$ of the field.

Around $z=0$, the field of the Gaussian takes the following simplified expression:

 \begin{equation} \label{eq:gaussian_perturbation}
\epsilon h_{x x}(\textbf{r},t) =  \epsilon  e^{ - \frac{( \textbf{n} \cdot \textbf{r})^2}{ \sigma^2 } + i [ \textbf{k} \cdot \textbf{r} - \omega t ] }
\end{equation}

with

$$\textbf{k} = k \widehat{\textbf{e}}_z\qquad \textbf{n} = \widehat{\textbf{e}}_x  $$

and $\epsilon$ representing a parameter that controls the strength of the perturbation. We can consider the perturbation field along the $\textbf{n}$ axis to be localized inside our region $\mathcal{R}$ of size $2 \sigma$ around the origin, and the points outside this region are essentially flat.

\section*{Results}

\begin{figure}[htp] \centering{
\includegraphics[scale=0.4]{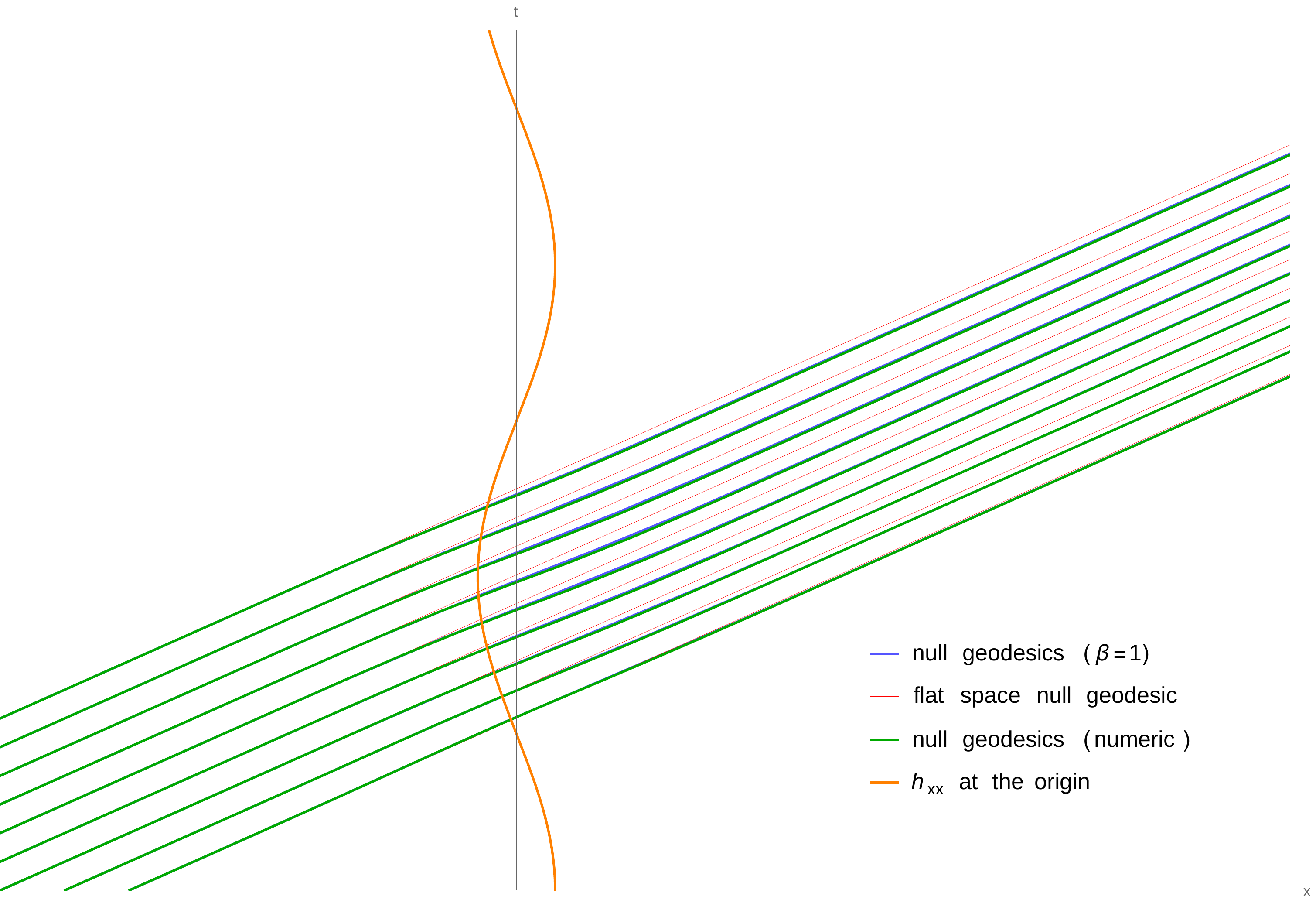}}
\caption{Null geodesics near the waveform valley for $\epsilon=0.6$}
\label{fig:null_geo_valley}
\end{figure}

\begin{figure}[htp] \centering{
\includegraphics[scale=0.4]{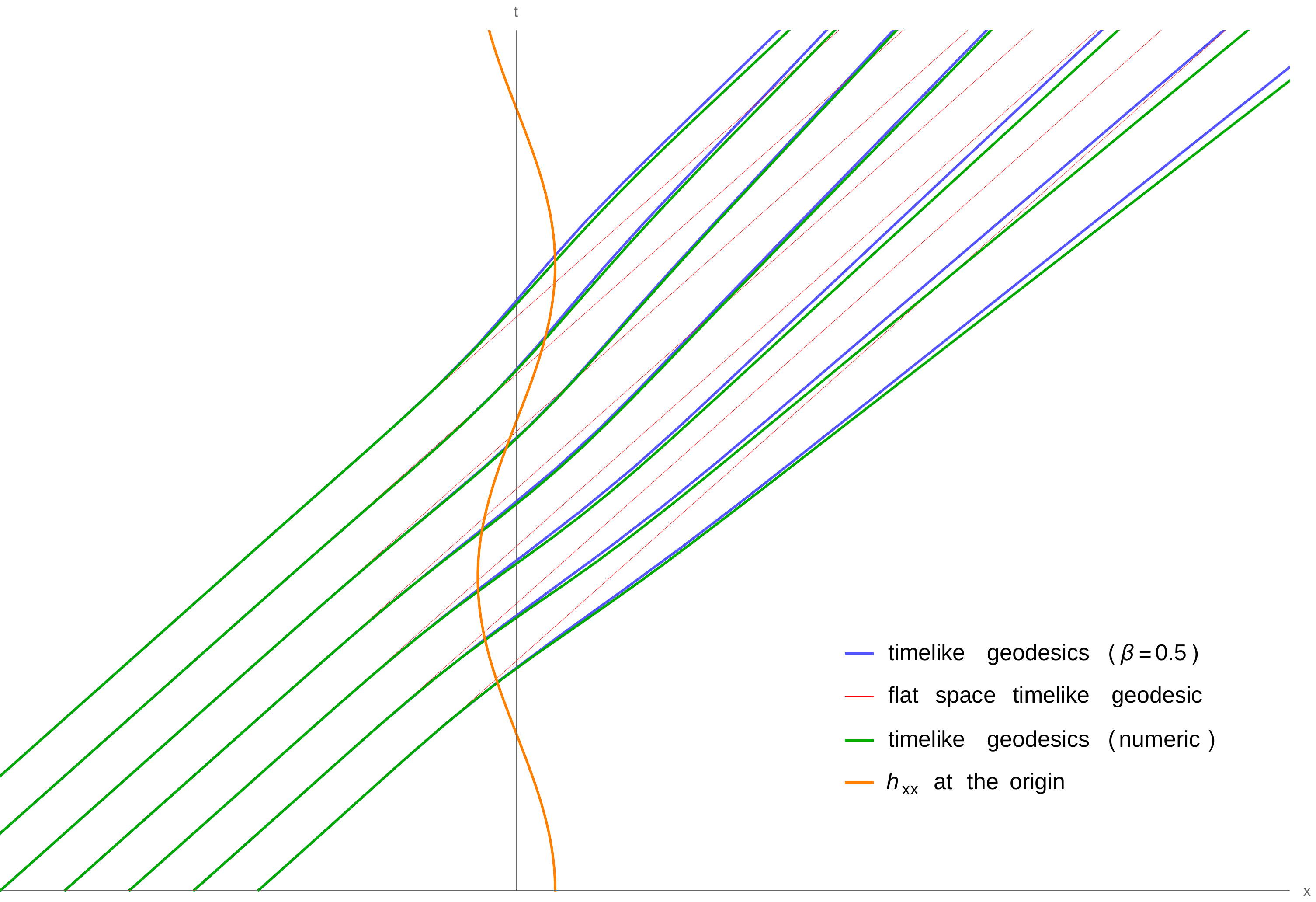}}
\caption{Timelike geodesics near the waveform valley for $\epsilon=0.6$}
\label{fig:timelike_geo_valley}
\end{figure}

Given the complex nonlinear geodesic equations that result from the Gau\-ssian perturbation eq.(\ref{eq:htt_transform}) and (\ref{eq:xx_component}), we were only able to compute analytically the first order correction to the geodesic null and time-like rays crossing the field, and its derivation is detailed in appendix \ref{ap:pert_geo}. We also used numerical integration on the system of geodetical differential equations (restricted to the subspace $z=y=0$) to confirm the validity of the approximation on the linearised regime ($\epsilon \ll 1$). However, the similarities of the numerical solutions with the analytical approximation for high values of $\epsilon$ are an important clue that suggest that the qualitative behavior of the first order approximation remains valid well into the nonlinear regime. Figures \ref{fig:null_geo_valley} and \ref{fig:timelike_geo_valley} compare first order corrections with the full numerical solutions for $\epsilon = 6 \times 10^{-1}$.

Fig.(\ref{fig:null_geo_valley}) shows the null geodesics propagation. We see that rays entering $\mathcal{R}$ while the metric components $g_{xx}$ are shrinking, end up \emph{advanced} in the asympto\-tic flat region, relative to their flat spacetime evolution. Advanced wavefronts will be perceived by distant observers as having arrived \emph{before} they would have when the whole spacetime was flat. The opposite happens when $g_{xx}$ starts increasing again; light pulses become delayed when they cross during the waxing phase.

For all practical purposes, advanced pulses will be effectively travelling faster than light inside $\mathcal{R}$, when compared with other geodesics in the rest of the flat Minkowski spacetime that do not cross the field region. This does not contradict the fact that these geodesics are always moving at or below the universal speed of light according to measurements performed by local free-falling observers inside or outside $\mathcal{R}$. However, distant observers will still measure that some geodesics cross the region in less time because spacetime has shrinked in a preferential direction for their whole transit inside $\mathcal{R}$.

In fig.(\ref{fig:timelike_geo_valley}) we can see the flow of time-like geodesics for $\beta=0.5$. Paths that enter the waning phase of the waveform early, are able to escape and accelerate from the field region. On the other hand, paths that enter the waveform late when the waxing phase begins, suffer deceleration.

\section*{Discussion}
In this work, we have established that within GR, certain gravitational waveforms can result in geodesics that arrive at distant points earlier than light signals in flat spacetime. We presented an example waveform that can be used to manifest FTL behavior, and obtained an analytic first order perturbative approximation of geodesics appro\-aching the field region. We notice that the timing of entrance to the field region determines the asymptotic delay or advance of signals. The optimal shortcut geodesics are those that cross the field region as close as possible to the time and position of the minimum of the metric waveform.

Since our analysis applies fundamentally to linearized theory, we have that small waveforms will result in small, but non-zero advancements (or delays) on geodesic congruences. However, comparison of the analytic perturbative approximation with numerical solutions of the system of geodetical equation (which were done restricted to $z=y=0$ to avoid computing the $h_{xz}$ component explicitly), suggest that this phenomena should extend into high fields and nonlinear theory.

Compared to other FTL schemes like the Alcubierre drive or Lorentzian wormholes, which rely on unphysical matter fields to stabilize the geometry, the current approach relies only on gravitational wave generation and transmission through empty space. Assuming the daunting problem of astronomical scale gravitational wave generation is somehow solved, this method could in principle enable FTL travel without appealing to exotic physics. However a detailed analysis of tidal forces is required before assessing the feasibility of this scheme for transit of payloads.

The nature of the shortcut generation involves the creation of waveforms that compress and dilate spacetime in the direction of flight. In order for signals (or ships) to be able to take advantage of the metric-contracting fields, they must carefully control their timing synchronization, in order to cross the field regions as close as possible to the compression valley, where the distance is minimal between opposing sides of the field region. The region must be crossed in substantially less than $T/2$, with $T$ being the period of the gravitational wave. Even if the compression of each field region is small, large distance reduction could be accomplished by bridging many pre-configured field regions in a timely manner. It is conceivable that other field configurations exist that achieve better distance compression patterns. Even without exploiting the FTL aspects of the field, time-like geodesics can still be substantially accelerated or decelerated with special field configurations of this type, while remaining in free fall during the transit.

Due to the transversal nature of gravitational waves, the gravitational sources must be distributed ortho\-gonally to the direction of desired FTL geodesic path. These gravitational beams have to be precisely oriented and timed decades in advance, as gravitational waves propagate at the speed of light. This implies some sort of deployment of a wide scale network of gravitational generators around entire star clusters. This presents an unfathomably hard logistic and technological problem. Perhaps, some future Type-III civilization, millions of years in the future, might manage to address them.

\appendix

\section{Appendix: Perturbative geodesic approximation}
\label{ap:pert_geo}

We want to compute $\delta x^{i}$ to first order in $\epsilon$ such that 

$$ \frac{d^2 (x_{[0])}^{\mu} + \delta x^{i})}{d \lambda^2} = - \Gamma^{\mu}_{\alpha \beta}(\eta + \epsilon h) \frac{d (x_{[0]}^{\alpha} + \delta x^{\alpha})}{d \lambda} \frac{d (x_{[0]}^{\beta} + \delta x^{\beta})}{d \lambda} $$

We write $\delta x$ as an expanded formal series of powers of $\epsilon$

$$ \delta x^{\mu} = \sum_i \delta x^{\mu}_{[i]} \epsilon^i $$

Our base geometry is flat geometry $\eta_{\mu \nu}$, and our base geodesic $x_{[0]}$ is just any flat geodesic ray that matches our desired initial conditions at $\lambda = 0$.

It is easy to see that the first order correction to our geodesic satisfies the equation:

$$\frac{d^2 \delta x^{\mu}_{[1]}}{d \lambda^2} = -  \Gamma^{\mu}_{\alpha \beta}(x_{[0]}(\lambda))_{[1]} \frac{d x_{[0]}^{\alpha}}{d \lambda} \frac{d x_{[0]}^{\beta} }{d \lambda}$$

Where the $_{[1]}$ suffix in the Christoffel symbol denotes first-order terms on $\epsilon$. Our first-order correction to the geodesic will be

$$ x^{\mu}(\lambda) = x_{[0]}^{\mu}(\lambda) - \epsilon \int_{0}^{\lambda}\int_{0}^{\lambda'} \Gamma^{\mu}_{\alpha \beta}(x_{[0]}(\hat{\lambda}))_{[1]} \frac{d x_{[0]}^{\alpha}}{d \hat{\lambda}} \frac{d x_{[0]}^{\beta} }{d \hat{\lambda}} d \hat{\lambda}  d \lambda'  $$

On our case 

$$x_{[0]}(\lambda)= ( \lambda, p_0 + c \beta \lambda , 0 , 0 ) $$

which describes a geodesic in the background geometry being sent in the $\widehat{\textbf{e}}_x$ direction at $t=\lambda=0$ and $x=p_0$. The background geodesic is a null ray for $\beta=1$.

We have that to first order, our $t$, $x$ and $y$ component integrals are given in terms of $h_{xx}$ only. The $z$ component depends on derivatives $h_{xz}$ as well, and we will not evaluate this component, as our main focus is the movement of geodesics in the $xt$ plane.

\begin{subequations}
\label{eq:geodesic_metric_components}
\begin{equation}
 - \Gamma^{t}_{\alpha \beta}(x_{[0]}(\lambda))_{[1]} \frac{d x_{[0]}^{\alpha}}{d \lambda} \frac{d x_{[0]}^{\beta} }{d \lambda} = -\frac{\beta^2 h_{+}}{2} \partial_t h_{xx} \Big |_{x_{[0]}(\lambda)}
\end{equation}

\begin{equation}
 - \Gamma^{x}_{\alpha \beta}(x_{[0]}(\lambda))_{[1]} \frac{d x_{[0]}^{\alpha}}{d \lambda} \frac{d x_{[0]}^{\beta} }{d \lambda} = -\frac{ c^2 \beta^2 h_{+}}{2} \partial_x h_{xx} \Big |_{x_{[0]}(\lambda)} - c \beta h_{+} \partial_t h_{xx} \Big |_{x_{[0]}(\lambda)}
\end{equation}

\begin{equation}
  - \Gamma^{y}_{\alpha \beta}(x_{[0]}(\lambda))_{[1]} \frac{d x_{[0]}^{\alpha}}{d \lambda} \frac{d x_{[0]}^{\beta} }{d \lambda} = -\frac{ c^2 \beta^2 h_{+}}{2} \partial_y h_{xx} \Big |_{x_{[0]}(\lambda)}
\end{equation}

\end{subequations}

The resulting integrals for the Gaussian geometry eq.(\ref{eq:xx_component}) are expressed in terms of the Error functions 

$$\text{erf}(z) = \frac{2}{\sqrt{\pi}} \int_0^z e^{-t^2} dt$$

The lengthy resulting expressions are given below:

\begin{subequations}
\begin{multline}
\delta x_{[1]}^{t}(\lambda) = 
\frac{I \sqrt{\pi} \beta \omega h_{+} }{4 c \sigma}  \exp{\Big ( - \frac{\omega( 4 c p_0 \beta I + \omega \sigma^2 )}{4 c^2 \beta^2} \Big )} \times \\ 
\Bigg \{ e^{\frac{2 I p_0 \omega }{\beta  c}} \left(\frac{\left(2 \beta  c p_0 +I \sigma ^2 \omega \right)}{2 \beta ^2 c^2} \left[ \text{erf}\left(\frac{\beta  c \lambda }{\sigma }+\frac{I \sigma  \omega }{2 \beta  c}+\frac{p_0}{\sigma }\right) -  \text{erf}\left(\frac{p_0}{\sigma }+\frac{I \sigma  \omega }{2 \beta  c}\right) \right] \right. \\
\left. + \frac{\sigma}{\sqrt{\pi } \beta  c} \left[ \exp \left(-\frac{\left(2 \beta  c (\beta  c \lambda +p_0)+I \sigma ^2 \omega \right)^2}{4 \beta ^2 c^2 \sigma ^2}\right)- e^{-\frac{\left(2 \beta  c p_0+I \sigma ^2 \omega \right)^2}{4 \beta ^2 c^2 \sigma ^2}} \right] + \lambda  \text{erf}\left(\frac{\beta  c \lambda +p_0}{\sigma }+\frac{I \sigma  \omega }{2 \beta  c}\right)\right) \\
   - \frac{\left(2 \beta  c p_0 - I \sigma ^2 \omega \right)}{2 \beta ^2 c^2} \left[ \text{erf}\left(\frac{\beta  c \lambda }{\sigma }-\frac{I \sigma  \omega }{2 \beta c}+\frac{p_0}{\sigma }\right) - \text{erf}\left(\frac{p_0}{\sigma }-\frac{I \sigma  \omega }{2 \beta  c}\right) \right] \\
   - \frac{\sigma}{\sqrt{\pi } \beta  c} \left[ \exp \left(-\frac{\left(2 \beta  c (\beta  c \lambda +p_0)-I \sigma ^2 \omega \right)^2}{4 \beta ^2 c^2 \sigma ^2}\right) - e^{-\frac{\left(2 \beta  c p_0 - I \sigma ^2 \omega \right)^2}{4 \beta ^2 c^2 \sigma ^2}} \right] + \lambda \text{erf}\left(\frac{p_0}{\sigma }-\frac{I \sigma  \omega }{2 \beta  c}\right) \\
   - \lambda  e^{\frac{2 I p_0 \omega }{\beta  c}}
   \text{erf}\left(\frac{p_0}{\sigma }+\frac{I \sigma  \omega }{2 \beta  c}\right) - \lambda  \text{erf}\left(\frac{\beta  c \lambda
   +p_0}{\sigma }-\frac{I \sigma  \omega }{2 \beta  c}\right)  \Bigg \}
\end{multline}
\begin{multline}
 \delta x_{[1]}^{x}(\lambda) = \frac{\beta  c \lambda  h_{+} }{\sigma ^2} e^{-\frac{p_0^2}{\sigma ^2}}+\frac{\sqrt{\pi } h_{+} }{4
   \sigma } e^{-\frac{\omega  \left(\sigma ^2 \omega +4 I \beta  c p_0\right)}{4 \beta ^2 c^2}} \left(\text{erf}\left(\frac{p_0}{\sigma }-\frac{I \sigma  \omega }{2 \beta c}\right)-\text{erf}\left(\frac{\beta  c \lambda }{\sigma }-\frac{I \sigma  \omega }{2 \beta  c}+\frac{p_0}{\sigma }\right)\right) \\
   -\frac{I \sqrt{\pi } \lambda  \omega  h_{+} }{4 \sigma } e^{\frac{\omega  \left(-\sigma ^2 \omega +4 I \beta  c p_0\right)}{4 \beta ^2 c^2}}
   \text{erf}\left(\frac{p_0}{\sigma }+\frac{I \sigma  \omega }{2 \beta  c}\right) + \frac{\sqrt{\pi } \lambda  \omega  h_{+} }{4 \sigma } e^{-\frac{\omega  \left(\sigma ^2 \omega +4 I \beta  c p_0\right)}{4 \beta ^2 c^2}} \text{erfi}\left(\frac{\frac{\sigma ^2 \omega }{\beta  c}+2 I p_0}{2 \sigma }\right) \\
   + \frac{\sqrt{\pi } h_{+} }{4 \sigma } e^{-\frac{\omega  \left(\sigma ^2 \omega - 4 I \beta  c p_0\right)}{4 \beta ^2 c^2}}
   \left(\text{erf}\left(\frac{p_0}{\sigma }+\frac{I \sigma  \omega }{2 \beta  c}\right)-\text{erf}\left(\frac{\beta  c \lambda }{\sigma }+\frac{I \sigma  \omega }{2 \beta  c}+\frac{p_0}{\sigma }\right)\right) \\
   + \frac{I \sqrt{\pi } \omega  h_{+} }{4 \sigma } e^{\frac{\omega  \left(-\sigma ^2 \omega +4 I \beta  c p_0\right)}{4 \beta ^2 c^2}} \left[ \left(\frac{I \sigma ^2 \omega }{2 \beta ^2 c^2}+\frac{p_0}{\beta  c}+\lambda \right) \text{erf}\left(\frac{\beta  c \lambda }{\sigma }+\frac{I \sigma \omega }{2 \beta  c}+\frac{p_0}{\sigma }\right) \right. \\
   \left. -\frac{\left(2 \beta  c p_0+I \sigma ^2 \omega \right)}{2 \beta ^2 c^2}
   \text{erf}\left(\frac{p_0}{\sigma }+\frac{I \sigma  \omega }{2 \beta  c}\right)+\frac{\sigma }{\sqrt{\pi } \beta 
   c} \left(\exp \left(-\frac{\left(2 \beta ^2 c^2 \lambda +2 \beta  c p_0+I \sigma ^2 \omega \right)^2}{4 \beta ^2 c^2 \sigma
   ^2}\right)-e^{-\frac{\left(2 \beta  c p_0+I \sigma ^2 \omega \right)^2}{4 \beta ^2 c^2 \sigma ^2}}\right) \right] \\
   -\frac{\omega h_{+} }{8 \beta ^2 c^2 \sigma } e^{-\frac{\omega  \left(\sigma ^2 \omega +4 I \beta  c p_0\right)}{4 \beta ^2 c^2}} \left[\sqrt{\pi } \left(2 \beta ^2 c^2 \lambda +2 \beta  c p_0-I \sigma ^2 \omega \right) \text{erfi}\left(\frac{2 I \beta  c \lambda +\frac{\sigma ^2 \omega }{\beta  c}+2 I p_0}{2 \sigma }\right) \right. \\
   \left. + \sqrt{\pi } \left(-2 \beta  c p_0+I \sigma ^2 \omega \right) \text{erfi}\left(\frac{\frac{\sigma ^2 \omega
   }{\beta  c}+2 I p_0}{2 \sigma }\right)-2 I \beta  c \sigma  \left(e^{\frac{\left(\frac{\sigma ^2 \omega }{\beta  c}+2 I
   p_0\right)^2}{4 \sigma ^2}}-e^{\frac{\left(2 I \beta  c \lambda +\frac{\sigma ^2 \omega }{\beta  c}+2 I p_0\right)^2}{4 \sigma
   ^2}}\right)\right]
\end{multline}
\end{subequations}

Despite the presence of complex arguments, the expressions remains real if the parameters are real.

\bibliography{main}

\begin{thebibliography}{1}

\bibitem{Alcubierre:1994tu}
Miguel Alcubierre.
\newblock {The Warp drive: Hyperfast travel within general relativity}.
\newblock {\em Class. Quant. Grav.}, 11:L73--L77, 1994.

\bibitem{Coule_No_warp_drive_98}
D~H Coule.
\newblock No warp drive.
\newblock {\em Classical and Quantum Gravity}, 15(8):2523, 1998.

\bibitem{Finazzi:2009jb}
Stefano Finazzi, Stefano Liberati, and Carlos Barcelo.
\newblock {Semiclassical instability of dynamical warp drives}.
\newblock {\em Phys. Rev.}, D79:124017, 2009.

\bibitem{Ford:1994bj}
L.~H. Ford and Thomas~A. Roman.
\newblock {Averaged energy conditions and quantum inequalities}.
\newblock {\em Phys. Rev.}, D51:4277--4286, 1995.

\bibitem{Krasnikov:1995ad}
S.~V. Krasnikov.
\newblock {Hyperfast travel in general relativity}.
\newblock {\em Phys. Rev.}, D57:4760--4766, 1998.

\bibitem{Natario:2001tk}
Jose Natario.
\newblock {Warp drive with zero expansion}.
\newblock {\em Class. Quant. Grav.}, 19:1157--1166, 2002.

\end{thebibliography}
\bibliographystyle{plain}

\end{document}